\documentclass[a4paper,11pt]{article}
\usepackage{pos}

\title{Multigrid Solver on Fugaku}
\author[a]{Ken-Ichi Ishikawa}
\author*[b]{Issaku Kanamori}
\author[c,d]{Hideo Matsufuru}

\affiliation[a]{%
Core of Research for the Energetic Universe,\\
Graduate School of Advanced Science and Engineering, Hiroshima University,\\
1-3-1 Kagamiyama, Higashi-Hiroshima 739-8526, Japan}

\affiliation[b]{%
 RIKEN Center for Computational Science (R-CCS), \\
 7-1-26, Minatojima Minamimachi, Kobe 650-0047, Japan}

\affiliation[c]{Computing Research Center, High Energy Accelerator Research Organization (KEK),\\
 1-1 Oho, Tsukuba 305-0801, Japan}

\affiliation[d]{School of High Energy Accelerator Science, Graduate University for Advanced Studied (SOKENDAI),\\
 1-1 Oho, Tsukuba 305-0801, Japan}

\emailAdd{ishikawa@theo.phys.sci.hiroshima-u.ac.jp}
\emailAdd{kanamori-i@riken.jp}
\emailAdd{hideo.matsufuru@kek.jp}

\abstract{
We report an implementation of a multigrid solver for the Clover fermion
on supercomputer Fugaku,
which uses A64FX CPU with Arm architecture.
On Fugaku, a highly optimized implementation of BiCGStab solver with domain decomposed
preconditioner for the Clover fermion, called QCD Wide SIMD library (QWS),
is available.
We use the preconditioner in QWS as a smoother of the multigrid solver.
Our implementation shows reasonably good scaling and performance.
The code is developed by using Bridge++ code framework and its extension.
}

\FullConference{%
 The 38th International Symposium on Lattice Field Theory, LATTICE2021
  26th-30th July, 2021
  Zoom/Gather@Massachusetts Institute of Technology
}

\begin{document}
\maketitle

\section{Introduction}

The most time-consuming part of typical lattice QCD simulations is
linear solvers to solve Dirac equations by using iterative methods.
As the quark mass becomes smaller, the number of iterations
needed to solve the system increases and thus we need efficient preconditioners.
The multigrid method is one of the most successful methods to accelerate
the solvers for lattice QCD~\cite{Babich:2010qb}.

In this article, we focus on the implementation of a multigrid solver
for supercomputer Fugaku.  Fugaku, the number 1 machine in the Top 500
list as of November 2021, adopts a new processor named A64FX 
with Armv8.2-A architecture with Fujitsu extension~\cite{a64fx}.
We usually need a specific tuning to each new architecture
to achieve high performance.
For lattice QCD application on Fugaku, 
we have developed a solver library for the Clover fermion,
called QCD Wide SIMD (QWS) library~\cite{Ishikawa:2021iqw, nakamuraLat2021}, as a part of 
the co-design activities for the development of Fugaku \cite{codesign}.
Not only being a library with high performance,
but QWS also provides us an example
of an efficient data layout for lattice QCD simulation on Fugaku.
QWS is a nested BiCGStab solver with a domain decomposed Schwartz
Alternating Procedure (SAP) preconditioner.
Its development has an aspect of a testbed for frequent neighboring communications and global reductions,
and it indeed achieved an ideal weak scaling up to more than 140,000 nodes~\cite{Ishikawa:2021iqw}.
Although it also gives good performance in realistic situations 
such as $O(100)$ nodes and almost physical light quark mass,
adopting better algorithms may further reduce the elapsed time.
Multigrid solvers are good candidates.
The purpose of this work is to provide a tuned version of the multigrid solver for Fugaku making use of the QWS as a part of implementation.

It is desirable, however, to minimize the machine-specific codes
to reduce the cost of the implementation as well as the maintenance. 
We successfully prepared a framework for multigrid solver using Bridge++ code set \cite{%
Kanamori:2021rwy% ICCSA2021
}, where only machine-specific fermion operators and some basic linear algebra routines are needed.
The multigrid solver contains several solvers: an outer solver, 
a smoother, and a coarse grid solver.
The outer solver and the smoother
work on the original fine grid lattice.  We need to prepare fine and coarse operators tuned to Fugaku.
  Following the strategy of DD-$\alpha$AMG 
\cite{Frommer:2013fsa},
we adopt multiplicative SAP for smoother, for which we make use of the implementation in QWS.
Although combining with QWS is mainly for a performance reason,
our second purpose is to provide an example of using QWS from an application code.
The source code of the multigrid solver will be publicly available 
as a part of Bridge++ 2.0~\cite{%
Akahoshi:2021gvk% CCP21
}.

In the next section, we provide the specification of Fugaku and introduce QWS.
We explain the details of our implementation in Sec.~\ref{sec:implementaion}.
We then provide the benchmark result in Sec.~\ref{sec:performance} before
giving the summary and outlooks in Sec.~\ref{sec:summary}.

\section{Supercomputer Fugaku and QWS}
\label{sec:fugaku}
The supercomputer Fugaku is installed at RIKEN Center for Computational Science
in Kobe, Japan and has started shared service in March 2021.
Its theoretical peak performance is 488 PFlops in double precision and 977 PFlops in single precision. 
Each node consists of one A64FX processor~\cite{a64fx}.
The processor has 48 cores that are grouped as 4 core memory groups (CMGs) for computation, and also has 2 or 4 assistant cores.
The SIMD, called scalable vector extension (SVE), is 512-bit wide.
The theoretical peak performance of one processor is 3,072 GFlops in double precision
at 2.0GHz operation, which is roughly the same as Intel Xeon Phi Knights Landing (KNL).
Each processor has an on-chip 32 GiB HBM2 memory whose bandwidth is 1,024 GB/s, which roughly amounts twice the KNL case.
The network is Tofu Interconnect D. Its injection bandwidth is 40.2 GB/s per node. 

The development of Fugaku adopted the co-design approach \cite{codesign},
where the hardware developer and software developer works cooperatively.
Lattice QCD is one of their 9 applications targeted in the co-design.
The QCD Wide SIMD (QWS) library (QWS) \cite{%
Ishikawa:2021iqw,%qws
nakamuraLat2021%
}
is one the outcomes.
It implements a nested BiCGstab solver with domain decomposed preconditioner 
for the Clover fermion and the inner solver achieved 102 PFlops
in single precision using 147,456 nodes,
 which is 10.2\% of the theoretical peak performance in the boost mode with 2.2GHz.

The inner solver in QWS is a single precision BiCGStab solver for the preconditioned operator $AM_{\mathrm{SAP}}$.
Here, $A$ is the Clover Dirac operator $D_{\mathrm{Clover}}$
with the Clover term preconditioned,
\begin{equation}
 A=C^{-1} D_{\mathrm{Clover}}
  = 1 +\kappa C^{-1}H,
\qquad C = 1+\frac{i}{2}\kappa c_{\mathrm{SW}}\sigma_{\mu\nu}F_{\mu\nu}
\end{equation}
with hopping parameter $\kappa$, 
the Clover term coefficient $c_{\mathrm{SW}}$,
and the hopping term $H$.
$M_{\mathrm{SAP}}$ is a domain decomposed multiplicative SAP preconditioner for $A$.
It uses 2 domains in each MPI process and a Jacobi iteration with a fixed number of iterations applied alternatively to each domain.
A single precision solution of $A x =b$ is constructed as
$x=M_{\mathrm{SAP}}y$ with $y$ obtained by solving $(AM_{\mathrm{SAP}})y =b$.
Combining with the Richardson-iteration, we obtain the solution of $Ax=b$ in
double precision.
To obtain the solution to the full operator $D_{\mathrm{Clover}}$,
the source vector $b$ should be preconditioned with the Clover term by multiplying $C^{-1}$.
In this way we can solve the full system $D_{\mathrm{Clover}}x =b$ by using QWS.
We can also use QWS to solve the SAP preconditioned system $AM_{\mathrm{SAP}}x=b$.
It also provides a routine for applying $M_{\mathrm{SAP}}$.
It is the user's responsibility to calculate $C$ and $C^{-1}$ from the gauge field
and to set the value of $C^{-1}$.

To minimize the communication latency, QWS uses a low level API
called uTofu for Remote Direct Memory Access (RDMA) together with a double buffering algorithm.
An important restriction is that due to the SIMD vector usage,
the local lattice in the $x$-direction must be a multiple of $32$ to make fully 
use of QWS.  We describe the details of the SIMD vector usage in the next section.

\section{Implementation Details}
\label{sec:implementaion}

We employ two-level multigrid process for simplicity.
We use single precision for the multigrid preconditioner.
The general structure of our implementation which does not use QWS is found in~\cite{Kanamori:2021rwy}.

The multigrid process contains the following four ingredients besides the setup process:
\begin{itemize}
 \item Restriction from the fine grid lattice to the coarse grid lattice
 \item Coarse grid solver
 \item Prolongation from the coarse grid to the fine grid lattice
 \item Smoother (fine grid solver)
\end{itemize}
The restriction and prolongation are domain decomposed algebraic processes
 with null space vectors.  We denote the number of null space vectors as $N_\mathrm{vec}$.
We adopt a BiCGStab solver for the coarse grid solver.
We use a post smoother after the prolongation and apply several (fixed numbers of) 
steps of multiplicative SAP.
For this, we call $M_{\mathrm{SAP}}$ of QWS from Bridge++.
Since the Dirac operator in QWS is preconditioned with the Clover term,
we need to apply $C^{-1}$ before applying $M_{\mathrm{SAP}}$, which is also available in QWS.
The outer solver is Flexible BiCGStab.
Before starting solving, we need the setup process to generate $N_{\mathrm{vec}}$ null space vectors.
The very initial null space vectors are generated by applying $M_{\mathrm{SAP}}$
 on $N_{\mathrm{vec}}$ random vectors and the initial coarse grid operator is generated by using them.
We then adaptively apply the multigrid preconditioner to them 4 times.
The coarse grid operator is also adaptively updated accordingly.
The combination of the domain decomposed adaptive multigrid with SAP smoother
is essentially the same as DD-$\alpha$AMG~\cite{Frommer:2013fsa},
 except for they did not use (flexible) BiCGStab
 but used (flexible/even-odd preconditioned) GMRes solvers.

Another difference from DD-$\alpha$AMG is that we use different domain sizes
for the coarse-graining process and SAP smoother.
The domain size for the restriction/prolongation is constrained by the SIMD vector usage as described below 
and we adopt $8\times 4 \times 4 \times 4$ lattice points, which divides each local lattice on one MPI process into $O(10)$--$O(100)$ domains.
For SAP, as we use $M_{\mathrm{SAP}}$ in QWS, we have only 2 domains in each local lattice.

\begin{figure}
 \center
 \includegraphics[width=0.4\linewidth]{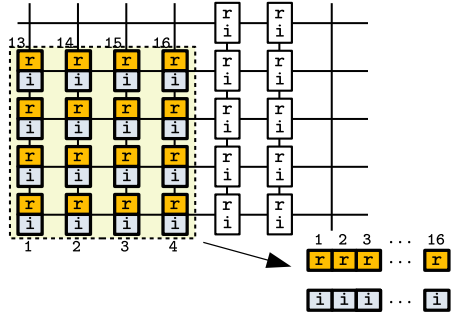}
 \caption{SIMD data layout.  We use the same layout for both fine and coarse grid systems.}
\label{fig:simd}

\end{figure}

The architecture-specific implementation appears in the usage of SIMD vector
most apparently.
During the development of QWS, it turned out that packing the real part and imaginary part of complex numbers into different SIMD vectors
is more efficient than packing them into the same vector.
As depicted in Fig.~\ref{fig:simd}, 
we pack the site degrees of freedom with two-dimensional tiling in 
the $x$-$y$ plane, where the complex numbers on $4\times 4$ lattice sites are packed into two SIMD vectors.
The same layout is used both on the fine grid lattice and the coarse grid lattice.
It is natural to constrain the size of the domain for the coarse-graining process 
to multiples of $4\times 4 \times 1 \times 1$ lattice and 
we adopt a domain that consists of $8\times 4 \times 4 \times 4$ lattice sites in the benchmark.
Since each domain gives one coarse lattice site,
this makes a typical number of coarse lattice sites divided by SIMD vectors
the same order of the number of threads, and thus we can still use thread parallelizations
for the site degrees of freedom.
A comment to higher-level of the multigrid extensions is in order.
In the next level, the number of lattice points divided by SIMD would be $O(1)$,
which is significantly smaller than the number of threads.
We can no longer use a simple thread parallelization for the site degrees of freedom and
we would need a different strategy of thread parallelization, \textit{e.g.}, parallelization for the null vectors.

QWS, from which we employ the SAP preconditioner $M_{\mathrm SAP}$ as a smoother on the fine grid, adopt a different packing scheme of the lattice sites.
It uses one-dimensional tiling in the $x$-direction which constrains the domain size 
for the SAP to be a multiple of 16 in the $x$-direction.
Each local lattice has two domains and they divide the local lattice in $x$-direction.
Therefore the local extent in $x$-direction must be a multiple of 32 for $M_{\mathrm{SAP}}$.  While it is acceptable to impose such a constraint on the fine grid lattice,
it is too restrictive on the coarse grid lattice.  One could use a different data
layout for the coarse grid lattice but we instead use the two-dimensional tiling
in Fig.~\ref{fig:simd} for both grids.
With this choice, we have a more flexible local lattice size, although the local lattice size we can combine with QWS is still restricted\footnote{%
Although the efficiency is lower, a SAP preconditioner without QWS is also available in Bridge++ and can be combined with the multigrid solver \cite{Kanamori:2021rwy}.}.
To bridge the data layouts between, suitable layout conversions
are inserted before and after the smoother.

We use MPI persistent communication with Fujitsu extension for the neighboring communications, except for those in $M_{\mathrm{SAP}}$ which directly calls uTofu API as stated in the previous section.
The Fujitsu extension internally utilizes the assistant cores to accelerate the communication
and helps to overlap between communication and computation.
We do not have to call \texttt{MPI\_Test} that some systems may require to trigger the communication in the background nor devote one thread for calling \texttt{MPI\_Wait}
during the overlapped computation.  Therefore all the computational cores fully work for the computation during the communication.

\section{Performance}
\label{sec:performance}

For benchmarking our implementation, we examine 3 configurations from the following sets (1 configuration from each):
\begin{itemize}
 \item A: $32^3\times 64$ lattice with $156$ MeV pion \cite{PACS-CS:2008bkb}.
 \item B: $64^3\times 64$ lattice with $512$ MeV pion \cite{Yamazaki:2012hi}.
 \item C: $96^3\times 96$ lattice with $145$ MeV pion \cite{Ishikawa:2015rho}.
\end{itemize}
As a reference implementation, we use a solver in LDDHMC \cite{PACS-CS:2008bkb}
tuned for Fugaku.
The tuned version is a reimplementation of QWS in domain decomposed HMC but 
employs different axis ordering from QWS, the $t$-direction loop is innermost and SIMDzed.
In measuring the performance, we allocate 1 MPI process per CMG, 
\textit{i.e.}, 4 MPI processes per node.  The compiler is the Fujitsu compiler with version 4.5.0.  We use the clang mode of the compiler for the multigrid solver and the trad mode for LDDHMC.

\begin{figure}
\center
\includegraphics[width=0.48\linewidth]{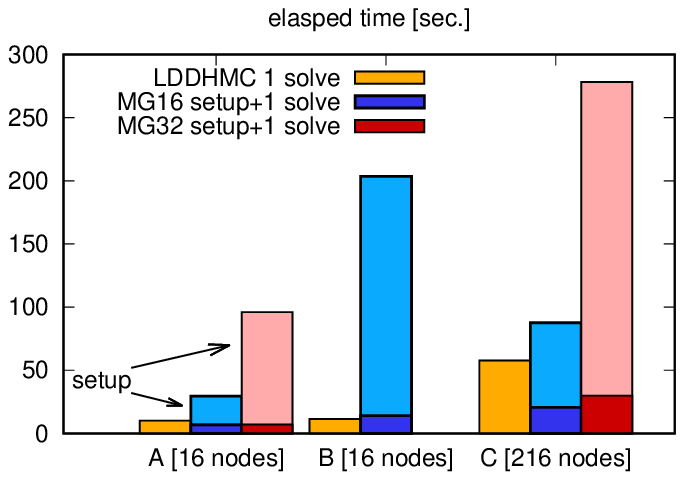}
\hfil
\includegraphics[width=0.48\linewidth]{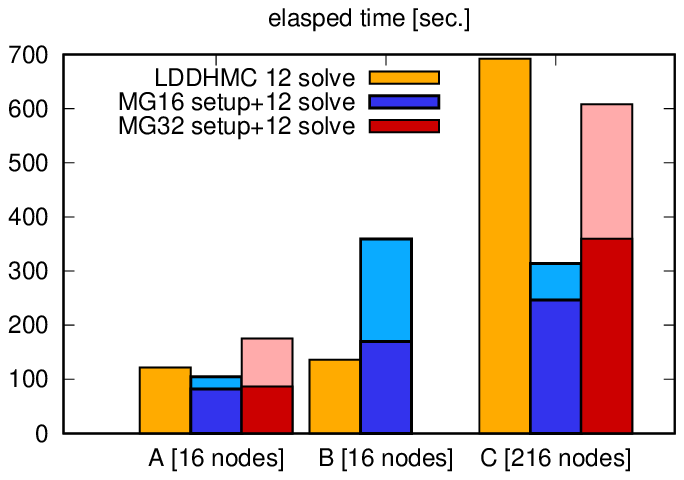}

\caption{Elapsed time to solve the linear equation for one source vector (left panel) and twelve source vectors (right panel).  MG16 and MG32 denote the multigrid solvers with $N_{\mathrm{vec}}=16$ and $32$, respectively.  MG32 for Set B on 16 nodes did not run because the memory size needed exceeds the available memory size.}
\label{fig:performance}

\end{figure}

Figure~\ref{fig:performance} shows the elapsed time to solve 1 equation (left panel) and 12 equations on the same configuration (right panel).
For the multigrid solvers, we examine two different numbers of null space vectors,
$N_{\mathrm{vec}}=16$ and $32$.
In the figure, the setup time, which is necessary to prepare the null space vectors and called only once for each configuration,
is also piled as a light color box to each elapsed time of the multigrid solver.
LDDHMC is always the fastest to solve one equation
because of the large overhead of the multigrid solver due to the setup process to generate null space vectors.
On the other hand, if we need to solve more equations with the same configuration, which often happens in the measurements, the overhead can be compensated by a shorter elapsed time in solving with the multigrid solver.
Especially for the set C whose quark mass is the lightest among the three sets, the time to solve
12 equations with multigrid solver ($N_{\mathrm{vec}}=16$) including the setup overhead is less than half the timing of LDDHMC.
On the other hand, for Set B, which corresponds to a rather heavy quark mass, 
the ``solve'' of the multigrid solver is slower than LDDHMC so the overhead 
cannot be compensated.

\begin{figure}
\center
\includegraphics[width=0.48\linewidth]{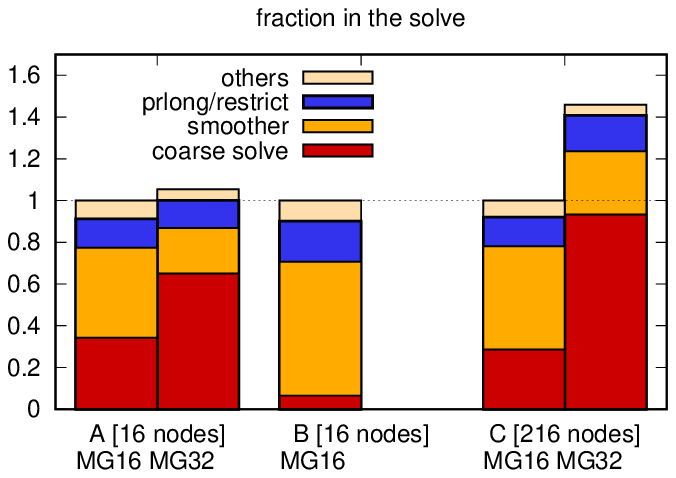}
\hfil
\includegraphics[width=0.48\linewidth]{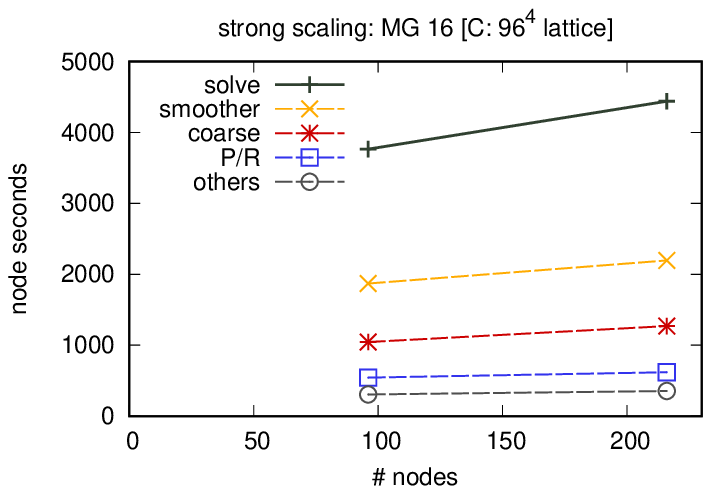}

\caption{Left panel: fraction of elapsed time for each component of the multigrid solver.
For each configuration, the elapsed time is normalized by the solving time without the setup with the $N_{\mathrm{vec}}=16$ case (MG16).
The result of MG32 for set B is missing because the required memory size 
exceeds the available system memory size.
Right panel: strong scaling of the multigrid solver with $N_{\mathrm{vec}}=16$ with set C.  P/R stands for the prolongation and restriction.} 
\label{fig:breakdown_and_scaling}

\end{figure}

The fraction of each component of the solving is shown in the left panel of
Fig.~\ref{fig:breakdown_and_scaling}.
It is natural to observe that having a larger number of null space vectors 
results in spending more time in the coarse grid solver.
Within the test cases, there is a tendency that spending more time in the smoother gives shorter elapsed time to solve the equation.
This is because the smoother from QWS is very efficient.
The performance of individual ingredients per node for set C with $N_{\mathrm{vec}}=16$ on 216 nodes are the following: 790 GFlops for the smoother, 91 GFlops for the coarse grid solver, 82 GFlops for the restriction, and 125 GFlops for the prolongation.
Note that the theoretical peak performance is 6,144 GFlops/node in single precision.

The right panel in Fig.~\ref{fig:breakdown_and_scaling} is a strong scaling plot
of the solving and each part, for set C with the $N_{\mathrm{vec}}=16$ case.
It is not a perfect scaling but a reasonably good scaling:
as the number of nodes increases by 2.25 times, the elapsed time reduces 
from 39.20 seconds to 20.56 seconds.
This corresponds to 86\% in the parallelization efficiency.
The scaling of the prolongation and restriction (P/R) is almost perfect.
The breaking of scaling mainly comes from the coarse solver and the smoother.

\section{Summary and Outlooks}
\label{sec:summary}

We implemented a multigrid solver for the Clover fermion on supercomputer Fugaku.
The code is implemented in the framework of Bridge++ incorporating a preconditioner from QWS, 
which is an outcome of the co-design activity for Fugaku.
Our implementation, therefore, provides an example of making an efficient use of QWS.
We adopted a data layout specific to the A64FX processor.  To use a SAP
preconditioner from QWS, the domain size for the SPA preconditioner is different
from the domain size for the coarse-graining process to make a coarse grid lattice.
For light enough quarks, the multigrid solver showed a shorter elapsed time
to solve one equation than a solver from LDDHMC, a reimplementation of QWS.
The overhead to prepare the null space vectors is not negligible 
so that the current implementation is not suitable to solely solve a single equation. If we need to solve several equations on the
same gauge configuration, the multigrid solver becomes more efficient in the total computational cost.  The performance largely owes to the high efficiency of the QWS.
The strong scaling is reasonably good.

For further improvement, we are planning to update the coarse solver and
the setup process. 
 Although the current two-level multigrid is practically enough,
 it would be interesting to investigate
 possible implementations for higher-level multigrid preconditioners.
The code will be available in the forthcoming release of Bridge++.

\subsection*{Acknowledgments}

We thank LQCD co-design team in flagship 2020 project Bridge++ project members and.  The configurations are from Japan Lattice Data Grid \cite{JLDG}.
The numerical simulation and code development were performed  on the supercomputer Fugaku at RIKEN Center for Computational Science and
 `Flow' at Information Technology Center, Nagoya University.
This work is supported by JSPS KAKENHI (%
JP19K03837,JP20K03961),
the MEXT as `Program for Promoting Researches on the Supercomputer
Fugaku' (Simulation for basic science: from fundamental laws of
particles to creation of nuclei) and
`Priority Issue 9 to be Tackled by Using the Post-K Computer' (Elucidation of the Fundamental Laws and Evolution of the Universe),
and Joint Institute for Computational Fundamental Science (JICFuS).

%\bibliographystyle{JHEP}
%\bibliography{lattice2021_proc.bib}

% from the bbl file

\providecommand{\href}[2]{#2}\begingroup\raggedright\endgroup

\end{document}